\documentclass[sigconf]{acmart}

\usepackage{booktabs} 

\setcopyright{rightsretained}
\copyrightyear{2021}
\acmYear{2021}
\acmConference{SA '21 Emerging Technologies:}{December 14-17, 2021}{Tokyo, Japan}

\setcitestyle{square}

\hypersetup{draft}

\begin{document}

\title{DroneStick: Flying Joystick as a Novel Type of Interface}

\author{Evgeny Tsykunov}
\affiliation{%
 \institution{Skoltech, Moscow, Russia}}
\email{Evgeny.Tsykunov@skoltech.ru}

 \author{Aleksey Fedoseev}
 \affiliation{%
  \institution{Skoltech, Moscow, Russia}}
\email{Aleksey.Fedoseev@skoltech.ru}

 \author{Ekaterina Dorzhieva}
 \affiliation{%
  \institution{Skoltech, Moscow, Russia}}
\email{Ekaterina.Dorzhieva@skoltech.ru}

\author{Ruslan Agishev}
\affiliation{%
 \institution{Skoltech, Moscow, Russia}}
\email{Ruslan.Agishev@skoltech.ru}

\author{Roman Ibrahimov}
\affiliation{%
 \institution{Skoltech, Moscow, Russia}}
\email{Roman.Ibrahimov@skoltech.ru}

\author{Derek Vasquez}
\affiliation{%
 \institution{Brigham Young University, Provo, Utah, US}}
\email{davasquez@fsu.edu}

\author{Luiza Labazanova}
\affiliation{%
 \institution{Skoltech, Moscow, Russia}}
\email{Luiza.Labazanova@skoltech.ru}

\author{Dzmitry Tsetserukou}
\affiliation{%
 \institution{Skoltech, Moscow, Russia}}
\email{D.Tsetserukou@skoltech.ru}

\renewcommand{\shortauthors}
{Tsykunov et al.}

\begin{abstract}
DroneStick is a novel hands-free method for smooth interaction between a human and a robotic system via one of its agents, without training and any additional handheld or wearable device or infrastructure. A flying joystick (DroneStick), being a part of a multi-robot system, is composed of a flying drone and coiled wire with a vibration motor. By pulling on the coiled wire, the operator commands certain motions of the follower robotic system. The DroneStick system does not require the user to carry any equipment before or after performing the required interaction. DroneStick provides useful feedback to the operator in the form of force transferred through the wire, translation/rotation of the flying joystick, and motor vibrations at the fingertips. Feedback allows users to interact with different forms of robotic systems intuitively. A potential application can enhance an automated `last mile' delivery when a recipient needs to guide a delivery drone/robot gently to a spot where a parcel has to be dropped. 
\end{abstract}



\begin{teaserfigure}
 \centering
 \includegraphics[width=\textwidth]{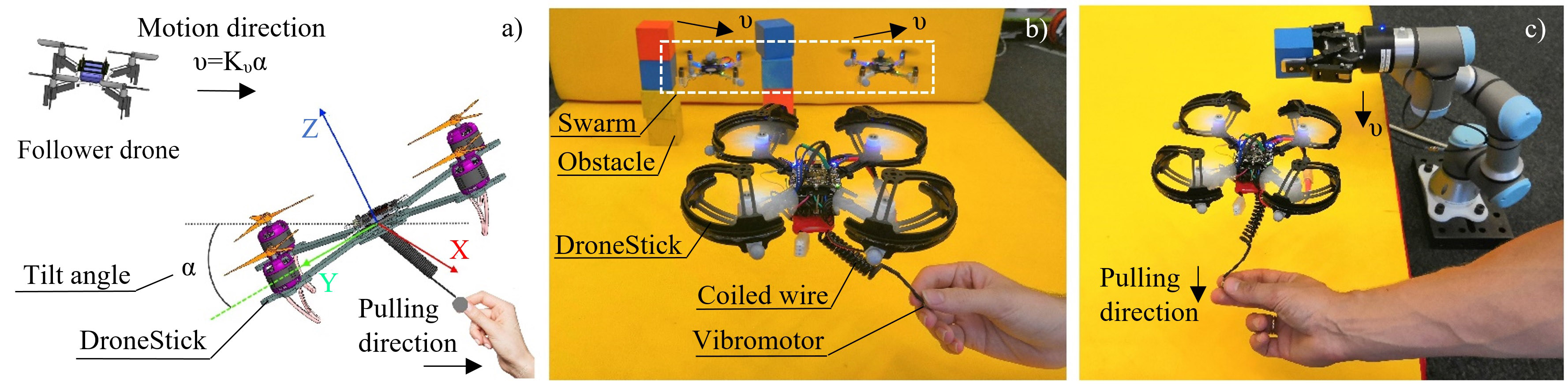}
 \caption{a) DroneStick principal of operation. b) Control of swarm of drones. c) Control of a robotic arm.}
 \label{first}
\end{teaserfigure}

\maketitle

\section{Introduction}

\subsection{Related work}
Productive interaction between a human and a robotic system often requires significant investment in training and equipment.
Direct control of teleoperated devices and machines is mainly accomplished by using joysticks, which help users successfully achieve guidance in remote areas.
In
\cite{Zhou_2018}, a swarm of 3-5 quadrotors was teleoperated by a single human operator using a desktop gaming joystick, and each quadrotor in the single-body swarm maintained a given state while avoiding collisions. Desktop joystick provides rich functionality, but its main disadvantages are that it is a desktop device and sometimes it is required additional training for non-expert users.
Different wearable displays \cite{Tsykunov18} or handheld controllers provide more portability but require the operator to either hold or wear the device during operation, in addition, to carry it before and after the interaction, which limits hands-free scenarios.

In the gesture-based interface proposed by
\cite{Podevijn_2014}, a swarm of ground robots receives commands from a human operator in the form of gestures, which are captured by an RGB-D Kinect camera. This approach allows operators to control robots without additional devices yet suffers from low accuracy and lack of feedback, limiting its potential applications.

In BitDrones proposed by \cite{Gomes_2016}, the UAV swarm was utilized as a self-levitating tangible display. The human-drone interaction for teleoperation was proposed in GridDrones by \cite{Braley_2018}. Authors suggested two types of tangible manipulations with a leader drone: uni-manual (slight pushing) touch and bi-manual (holding and moving) touch.
For the human, however, it can be hard to smoothly touch and move the leader drone, which leads to the flight stability issues. In addition, such disturbances propagate to the controlled robots, which leads to unintended rapid accelerations of the operated subsystem. After the interaction and during the release, the leader drone can receive an unpredictable initial impulse which requires compensation from a controller. Finally, during the bimanual interaction, the leader drone loses its degrees of freedom and is not able to translate or rotate (which can be used as a source of additional information).

\subsection{Main contribution}
In this paper, we introduce a novel Human-Robot Interaction (HRI) concept of using a single drone, DroneStick, which belongs to a larger robotic system, to intuitively control the other heterogeneous robotic agents in this system.
The main novelty of the DroneStick is that we propose to interact with the robotic system via one of its agents, without any additional handheld or wearable device or infrastructure. In comparison to
\cite{Zhou_2018, Podevijn_2014}
the user does not need to have a separate joystick, and there is no reason to set up an RGB-D camera on the ground or onboard.

DroneStick is a drone with an elastic wire (with power and control cables inside) and a handgrip with an embedded vibromotor attached to the bottom. When human approaches the robot to be guided (e.g., mobile, flying, articulated robot), DroneStick takes off from the launchpad and flies up to the operator. By pulling the grip with fingers, user commands the rest of the robots follow the direction of pulling.
Since DroneStick is a flying robot, one of its main advantages is that it can fly to a location that is convenient for the specific application.


In contrast to \cite{Braley_2018}, the generated control input is smooth by its nature due to the elastic hardware element in the interface: coiled wire and a small displacement of the drone.
In addition, we propose to use a vibromotor attached to the end of the wire to deliver tactile feedback to the operator (short alarm impulses). Tactile feedback can be used to inform operator of obstacle proximity or the state of power supply.
Thus, feedback from DroneStick is one of the main contributions and feature differentiating it from GridDrones project.

\section{DroneStick Technology}
DroneStick is a quadrotor that maintains its position during its use as a joystick. 
A coiled wire with a vibromotor at the end is connected to the bottom of the DroneStick, as shown in Fig. \ref{first}. 
While the operator is pulling the handgrip, the DroneStick attempts to maintain its spatial position, which causes a slight change in orientation and position.
Four parameters of the state of the DroneStick ($pitch, roll, z, yaw$) are used to calculate four control inputs (to control 4 D.o.F) for the follower robots. In case of the quadrotors it can be written as in Eq. \ref{eq:control_input}:

\begin{equation} \label{eq:control_input}
    \begin{bmatrix} V_{x} \\ V_{y} \\ V_{z} \\ \alpha \end{bmatrix}
    = K_{v}
    \begin{bmatrix}
    -pitch \\
    roll \\
    z - z_{d} \\
    yaw - yaw_{d}
    \end{bmatrix}
    *
    \begin{bmatrix}
    abs(pitch)>angle_{lim} \\
    abs(roll)>angle_{lim} \\
    abs(z - z_{d})>z_{lim} \\
    abs(yaw - yaw_{d})>yaw_{lim} 
    \end{bmatrix}
     \end{equation}
\newline     
where $V_{x}, V_{y}, V_{z}$ are the global velocities for the follower robots, $\alpha$ is the yaw angle of the follower robots, $K_{v}$ is the vector of scaling coefficients which determines the control sensitivity, and  $param_{lim}$ are the threshold values that filter noise.

The top priority from HRI perspective is to make the technology safe to operate. All propellers of DroneStick are covered with guards to protect against injury.
In addition, software safeguards have been implemented to prevent the follower robots from moving too quickly or uncontrollably.
When the DroneStick experiences extreme values in the state parameters (e.g., pitch, roll), the motors are shut down. The operator can use this functionality as an emergency stop. When the motors of the DroneStick shut down, all follower robots enter safety mode (e.g., safe landing).


\section{Conclusion}
The initial purpose of DroneStick was to intuitively control a swarm of drones with one of its agents. However, we propose to generalize the DroneStick technology to control heterogeneous robots, such as mobile, quadruped, swimming, flying robots and robotic arms. Thus, the proposed concept demonstrates the potential impact on a broad set of technologies changing the way we interact with machines.
DroneStick provides various forms of feedback to the operator to accommodate more intuitive and precise control of a robotic system.
Safety and flight stability of the DroneStick can be potential limitations.
Future work will be devoted to the evaluation of the technology by users.

\section{Acknowledgements}

The reported study was partly funded by RFBR and CNRS according to the research project No. 21-58-15006 and Skoltech-TUM grant.



\bibliographystyle{ACM-Reference-Format}
\bibliography{aaatemplate}
\end{document}